\documentclass[twocolumn,a4paper,prl,amssymb,nofootinbib,preprintnumbers]{revtex4}
\usepackage{epsfig}
\begin{document}
\title{``Phase Transitions in the Early Universe''\\
and\\
``Defect Formation''
\footnote{Lectures given at
a COSLAB Workshop on
``Cosmological Phase Transitions and Topological Defects'',
22--24 May 2003, Porto, Portugal.
}}

\author{Arttu Rajantie}

\affiliation{DAMTP, CMS, University of Cambridge,
Wilberforce Road, Cambridge CB3 0WA, United Kingdom
}

\begin{abstract}
In the first one of these two lectures, I 
give an introductory review of phase transitions 
in finite temperature field
theories.
I highlight the differences between theories with global and local
symmetries, and the similarities between cosmological phase transitions
and phase transitions in superconductors.
In the second one,
I review the theory of defect formation in finite temperature
phase transitions, emphasising
the differences between
systems with broken global and local symmetries.
These same principles apply to relativistic and condensed matter systems 
alike.
\end{abstract}

\preprint{DAMTP-2003-129}

\date{November 20, 2003}

\maketitle

\begin{center}
\bf\large 
Lecture I:\\
Phase Transitions in the Early Universe
\end{center}

\section{Introduction}

If one calculates the equilibrium properties of the Standard Model at
high temperatures, one finds several different 
phases~\cite{Kapusta,Linde,Bailin}. Below 100 MeV,
the elementary particles behave more or less as at zero
temperature. If we heat them up above this temperature, 
there is a transition to a
deconfined
phase in which quarks and gluons are not confined into hadrons.
At 100 GeV, there is another transition at which the weak and
electromagnetic interactions unify. Extensions of the Standard Model,
such as Grand Unified Theories (GUTs), typically predict new
phases at even higher temperatures.

This phase structure is important in cosmology~\cite{Kolb,Linde}
because the temperature of the universe was initially very high.
It seems likely that had to be higher than 100 GeV, because
otherwise
it is extremely difficult to explain the baryon asymmetry of the
universe --- the fact that there is so much matter in the universe, but
practically no antimatter.
As the universe expanded and cooled down to the
current 2.7 K, it must have therefore undergone at least the
deconfinement and electroweak phase transitions mentioned
above. Typical inflationary models predict much higher temperatures,
usually at least $10^{10}$ GeV,
and it is therefore likely that there were also other phase
transitions.

The cosmological consequences of these phase transitions depend
sensitively on their properties. In many cases the consequences
would be disastrous, and one can then rule out the theory that 
predicts the transition, or at least constrain its parameters.
Examples of this are grand unified theories with strongly
first-order phase transitions, as they would lead to a period of
``old inflation'', and to a very inhomogeneous universe, or theories
that predict massive domain walls or too many magnetic monopoles. 
In other cases, the effects of the phase transition may be subtle,
only barely observable, such as gravitational waves, or even
completely unobservable. Or, it may be that the consequences are
absolutely crucial for our own existence, as in the theory of
electroweak baryogenesis~\cite{Shaposhnikov}, 
which uses the electroweak phase transition
to explain the baryon asymmetry.

In this lecture, I will concentrate on those phase transitions that
can be understood as a spontaneous breakdown of either a global or a
local symmetry. This covers the electroweak and GUT transitions, and
to a certain extent also the deconfinement transition.
I will also point out similarities to phase transitions in condensed
matter systems such as superfluids and superconductors.
Unlike cosmological phase transitions, they can be studied experimentally
and can therefore be used to test our theoretical understanding of
phase transitions in quantum field theories.

\section{Definitions}

The properties of a thermodynamic system in thermal equilibrium are
given by the partition function $Z$ (see Ref.~\cite{Kapusta}), 
which can be generally written as
\begin{equation}
Z={\rm Tr}\ \hat\rho={\rm Tr}\ e^{-\beta \hat H},
\end{equation}
where $\hat H$ is the Hamiltonian of the system and $\beta=1/kT$ is the
inverse temperature.

In many cases, it is more
convenient to use its logarithm, which defines the free energy
$F=-T\ln Z$. This quantity has the property that its value is minimised in
thermal equilibrium.

Neither the partition function nor the free energy are observable
quantities. However, any observable can be expressed as the derivative
of $F$ with respect to some external field. For instance, the
expectation
value of the operator $\hat X$ is defined as
\begin{equation}
\langle \hat X\rangle = Z^{-1}{\rm Tr}\ \hat X\hat \rho,
\end{equation}
and if we define an external field $J_X$ by
\begin{equation}
Z(J_X)={\rm Tr}\ e^{-\beta \hat H + J_X\hat X},
\end{equation}
we find
\begin{equation}
\langle \hat X\rangle = \frac{\partial}{\partial J_X}\ln Z=
-\frac{1}{T}\frac{\partial F}{\partial J_X}.
\end{equation}
A phase transition is defined as a point at which the
free energy (or,
equivalently, the partition function) is non-analytic. 
In general, it follows that at least 
some observables are non-analytic at the transition point and act
as order parameters.

The simplest type of phase
transitions are first-order transitions, in which some expectation
value, say $\langle \hat X\rangle$, is
discontinuous. An everyday example of this is the transition between
water and ice.
Let us assume that the transition takes place at temperature $T_c$.
Because of the discontinuity, the state of the system at $T_c$ depends
on which side the critical temperature was approached
from. Consequently, it is possible to have two phases coexisting at
$T_c$, which is exactly what happens when we put ice into a glass of
water. As the temperature is gradually changed through $T_c$, 
the discontinuous jump from one value of 
$\langle \hat X\rangle$ to the other does not necessarily take place
instantaneously at $T_c$. For instance, water can remain liquid in a
supercooled state at
temperatures below $0^\circ$C for a relatively long time, if it is 
pure enough and completely unperturbed. This happens in a meteorological
condition known as freezing rain, in which the supercooled water freezes
as soon as it hits the ground, causing hazardous driving conditions.
More generally, this phenomenon is known as metastability.

\begin{figure}
\epsfig{file=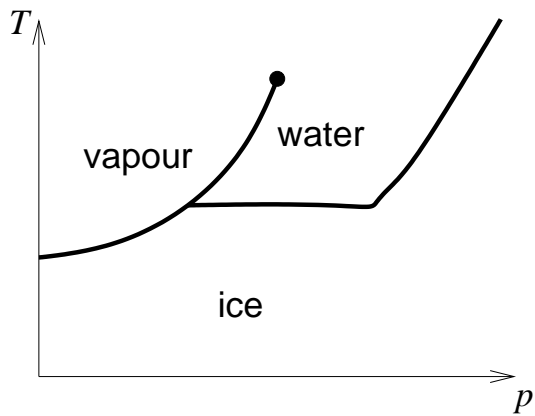,width=7cm}
\caption{
\label{fig:waterphases}
The phase diagram of water (greatly simplified). The lines are
first-order phase transitions. The transition line between water and
vapour ends at a second-order critical point.
}
\end{figure}

Second-order phase transitions are those in which every expectation
value is continuous, but some expectation value $\langle \hat
X\rangle$ has a discontinuous derivative. This is a
limiting case of a first-order transition in which the discontinuity
is taken to zero, and therefore it typically requires fine tuning of
some parameter. 
On a phase diagram, second order transitions are typically end points of
first order transition lines.
One example of  this is the
critical point of water at which the first-order transition line
between water and vapour ends. To reach this point, one has to tune
both the temperature and the pressure to
$T=647.096$ K and $p=22.064$ MPa. In contrast, reaching a first order
transition point, one only has to tune the temperature provided that
the pressure is below the critical value.

\begin{figure}
\epsfig{file=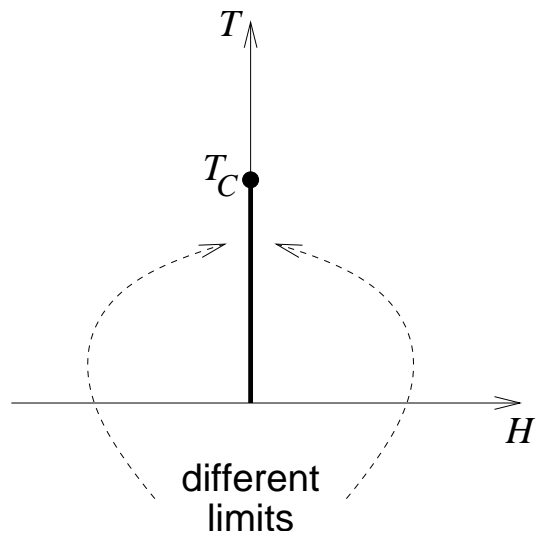,width=7cm}
\caption{
\label{fig:ferrophases}
The phase diagram of a typical ferromagnet. Because of the rotation
symmetry,
the line of
first-order transitions is on the $T$ axis. It ends in a second-order
transition point at the Curie temperature $T_C$.
The rotation symmetry is spontaneously broken at $T<T_C$.
}
\end{figure}

There is, however, an exception to this rule. There are cases in which
some symmetry forces the system to be on the first-order transition
line. This happens, for instance, in a ferromagnet. If the temperature
is below the Curie temperature, the magnetisation $\vec{M}$ becomes
non-zero even if there is no external magnetic field $\vec{H}$. 
Because the setting is
rotation invariant, no direction of $\vec{M}$ is
preferred. Therefore, the direction of $\vec{M}$, which  
depends on how the external field was taken to zero,
breaks the rotation invariance. This is known as spontaneous symmetry 
breaking.

If we imagine changing the external field $\vec{H}$ gradually through zero,
we see that $\vec{M}$ is discontinuous at $\vec{H}=0$. 
In other words, there is a first-order transition
at $\vec{H}=0$. If we increase the temperature keeping $\vec{H}$
zero, we are therefore guaranteed to reach the second-order end point
at the Curie temperature.

The above situation is very important in field theories, because 
they typically have a high degree of symmetry. As in the case of the
ferromagnet, it often happens at low temperatures that the state of
the system is not invariant under a symmetry of the theory and
this
typically leads to a second-order phase transition as the temperature
is increased.

\section{Global symmetry breaking in field theories}
As a simple example of a field theory with spontaneous symmetry 
breaking, let us consider the Lagrangian
\begin{equation}
{\cal L}=\frac{1}{2}\partial_\mu\phi\partial^\mu\phi
-\frac{\lambda}{4}\left(\phi^2-v^2\right)^2,
\label{equ:globLag}
\end{equation}
where $\phi$ is a real scalar field. The Lagrangian is symmetric under
$\phi\rightarrow -\phi$, which correspond to the group $\mathbb{Z}_2$.

The vacuum of the system is found by solving the Euler-Lagrange equations
for a constant field,
\begin{equation}
\frac{\partial{\cal L}}{\partial\phi}=0,
\label{equ:EL0}
\end{equation} 
which gives
\begin{equation}
\lambda\left(\phi^3-v^2\phi\right)=0.
\label{equ:ELexplicit}
\end{equation}
In addition to $\phi=0$, which is an unstable solution,
we find two solutions, $\phi=\pm v$, which are the 
possible vacuum states of the system. Neither choice is invariant
under the $\mathbb{Z}_2$ symmetry, and therefore the symmetry is spontaneously
broken in the vacuum.

At a non-zero temperature, the value of $\phi$ fluctuates from point to point. 
If these
thermal fluctuations are small, we can write the field as the sum of
the homogeneous background contribution $\phi_0$ and the fluctuation
$\delta\phi$, 
\begin{equation}
\phi(\vec{x})=\phi_0+\delta\phi(\vec{x}).
\end{equation}
We assume that these fluctuations are Gaussian, which is 
equivalent to the one-loop approximation in perturbation theory,
i.e., in the expansion
in terms of Feynman diagrams. In many cases, this is
a
good approximation. The homogeneous mode $\phi_0$ is chosen in such a
way that the mean value of the fluctuation in zero, 
$\langle\delta\phi\rangle=0$. The variance, however, is non-zero, and 
has the value~\cite{Kapusta}
\begin{equation}
\langle \delta\phi^2\rangle
=\int \frac{d^3p}{(2\pi)^3}
\frac{1}{\omega}\frac{1}{e^{\beta\omega}-1}
+ (\mbox{zero-temperature part}),
\label{equ:fluct0}
\end{equation}
where $\omega=(\vec{p}^2+m^2)^{1/2}$.
Here $m$ is the mass of the field in the vacuum around which the
expansion is done, and is given by the second derivative of the
potential,
\begin{equation}
m^2=\frac{\partial^2 V}{\partial \phi^2}.
\end{equation}
The zero-temperature part is actually divergent, but we do
not have to worry about that, because it is already absorbed in the
definition of the renormalised zero-temperature couplings.
For $T\gg m$, Eq.~(\ref{equ:fluct0}) can be expanded as
\begin{equation}
\langle \delta\phi^2\rangle
=\frac{T^2}{12} - \frac{mT}{4\pi}.
\label{equ:fluct}
\end{equation}

When the fluctuations $\delta\phi$ are taken into account, the 
homogeneous more $\phi_0$ feels an effective potential that is
different from the ``tree-level'' potential in
Eq.~(\ref{equ:globLag}).
This is easiest to see by averaging Eq.~(\ref{equ:ELexplicit}) over
fluctuations~\cite{Linde},
\begin{eqnarray}
0&=&
\lambda\left(
\langle\phi^3\rangle-v^2\langle\phi\rangle\right)
\nonumber\\
&=&
\lambda\left(\phi_0^3+
3\langle\delta\phi^2\rangle\phi_0-v^2\phi_0\right),
\end{eqnarray}
where we have used the fact that the fluctuations are symmetric
and odd moments must therefore vanish.
%
If we now use the leading term from Eq.~(\ref{equ:fluct}),
$\langle\delta\phi^2\rangle\approx T^2/12$, we find
\begin{equation}
\lambda\left(\phi_0^2+
\frac{T^2}{4}-v^2\right)\phi_0=0.
\end{equation}
This shows that the two non-zero minima only exist below
a critical temperature $T_c=2v$, and above that, the symmetry is
restored.

\begin{table*}
\begin{tabular}{c|c|c|c}
$T_c$ & theory & symmetry & order parameter\\
\hline
$10^{16}$ GeV & GUT & SU(5)$\rightarrow$
SU(3)$\times$SU(2)$\times$U(1) & $\Phi$, 24 real components\\
100 GeV & electroweak & $\rightarrow$ SU(3)$\times$U(1) &
$\phi$, 2 complex components\\
100 MeV & QCD & ``$\rightarrow$ U(1)'' & $\bar{q}q$ composite
\end{tabular}
\caption{
\label{table:ssb}
The phase transitions predicted by particle physics models.
}
\end{table*}

One can easily generalise this discussion to somewhat 
more complicated cases. If $\phi_i$ is a multicomponent real scalar field
with $i\in{1,\ldots,N}$, we can
consider the Lagrangian
\begin{equation}
{\cal L}=\frac{1}{2}\partial_\mu\phi_i\partial^\mu\phi_i
-\frac{\lambda}{4}\left(\phi_i\phi_i-v^2\right)^2,
\label{equ:multiLag}
\end{equation}
where summation over the indices $i$ is assumed.
This Lagrangian is invariant under SO($N$) rotations of the field $\phi_i$.
For two components, we can define a complex scalar field 
$\Phi=(\phi_1+i\phi_2)$, and we have
\begin{equation}
{\cal L}=\frac{1}{2}\partial_\mu\phi^*\partial^\mu\phi
-\frac{\lambda}{4}\left(\phi^*\phi-v^2\right)^2.
\label{equ:compLag}
\end{equation}
This Lagrangian is invariant under rotations of the complex phase angle of the 
scalar field $\phi$., which means that it has a U(1) symmetry.
Because of this symmetry, we can choose that the vacuum expectation 
value is real, and expand $\phi_1=\phi_0+\delta\phi_1$ 
and $\phi_2=\delta\phi_2$.
The variance of both $\delta\phi_1$ and $\delta\phi_2$ are
given by Eq.~(\ref{equ:fluct}).
The averaged Euler-Lagrange equation for the real part is
\begin{eqnarray}
0 &=&
\left\langle\frac{\partial{\cal L}}{\partial\phi_1}\right\rangle
\nonumber\\&=&
\lambda
\left(
\langle\phi_1^3\rangle
+
\langle\phi_2^2\phi_1\rangle - v^2\langle\phi_1\rangle
\right)
\nonumber\\&=&
\lambda\left(
\phi_0^2+3\langle\delta\phi_1^2\rangle
+\langle\delta\phi_2^2\rangle-v^2
\right)\phi_0.
\end{eqnarray}
Using Eq.~(\ref{equ:fluct}), we find
\begin{equation}
\lambda\left(\phi_0^2+
\frac{T^2}{3}-v^2\right)\phi_0=0,
\end{equation}
showing that this time the critical temperature is $T_c=\sqrt{3}v$,
but otherwise the conclusions are unchanged.

The restoration of broken symmetries at high temperatures 
is important for the early universe, because the universe was initially
extremely hot and because the theories of particle physics contain
several symmetries that are spontaneously broken at zero temperature. 
As shown in
Table~\ref{table:ssb}, these are associated with the hypothetical
Grand Unified Theory (GUT), the electroweak theory and quantum
chromodynamics (QCD). 

The QCD deconfinement 
phase transition takes place at around 100 MeV, and is the
only one of the three transitions that can be studied
experimentally. The transition cannot be fully understood within
perturbation theory, and it can be characterised in two different
ways, either as a transition from a confining to a deconfined phase,
or as spontaneous breakdown of the chiral symmetry.
We will adopt the latter point
of view.

If the up and down quarks were massless, the left and right handed quarks
would decouple. One could therefore carry out independent rotations between
the up and down quarks of the given handedness. In the QCD vacuum, this 
symmetry is spontaneously broken by the quark condensate 
$\langle \bar{q} q\rangle$, which shows how the left and right handed quarks
are aligned in the flavour space. Just like in a ferromagnet, one
would expect the symmetry to be restored at high temperatures, and this
seems to imply a second-order phase transition. 

However, the quarks are not actually massless and therefore the chiral 
symmetry is not exact. The quark mass term plays the same role as the magnetic 
field $\vec{H}$ in the ferromagnet phase diagram in Fig.~\ref{fig:ferrophases},
and shifts the system a little bit away from the first-order transition line.
Because there is no symmetry in the first place, there is no second order 
transition 
either. A first-order transition is still possible if the
phase diagram is more complicated than that in Fig.~\ref{fig:ferrophases}.

The GUT and electroweak phase transitions, on the other hand, 
involve exact symmetries, but local gauge symmetries instead of global ones.
Therefore, the above discussion does not apply them either, as we shall see
in the next section.

\section{Gauge symmetry breaking}

The Lagrangians discussed above had global symmetries. One has to carry out
the same rotation at every point in order for the Lagrangian to remain
unchanged. It is possible to modify the theory in such a way that it 
is invariant under position-dependent rotations as well. 
This is done by introducing
a gauge field $A_\mu$, and the local symmetries are therefore also known as
gauge symmetries. For the theory with a complex scalar in 
Eq.~\ref{equ:compLag}), the modified Lagrangian is
\begin{equation}
{\cal L}=
-\frac{1}{4}F_{\mu\nu}F^{\mu\nu}
+\frac{1}{2}D_\mu\phi^*D^\mu\phi
-\frac{\lambda}{4}\left(\phi^*\phi-v^2\right)^2,
\label{equ:gaugeLag}
\end{equation}
where we have used the covariant derivative $D_\mu=\partial_\mu+igA_\mu$
and the field strength 
tensor $F_{\mu\nu}=\partial_\mu A_\nu-\partial_\nu A_\mu$. 
This model is essentially a relativistic version of the Ginzburg-Landau
theory of superconductivity. Our results will therefore apply to 
superconductors on a qualitative level. The quantitative results differ,
though.

\begin{figure}
\epsfig{file=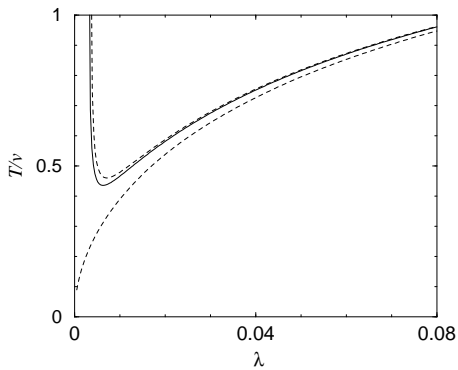,width=6cm}
\caption{
\label{fig:phasediag}
The critical temperature $T_c$ [solid line] in units of $v$ 
at $g=0.5$. The dashed lines show the metastability range.
}
\end{figure}

\begin{figure}
\epsfig{file=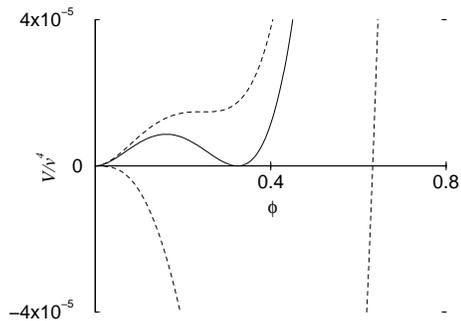,width=6cm}
\caption{
\label{fig:potential}
The effective potential at $T_c$ [solid line] for $g=0.5$, $\lambda=0.05$.
The dashed lines show the potential at the two ends of the
metastability range.
}
\end{figure}

The gauge coupling constant $g$ is simply the electric charge of the 
field $\phi$. (For superconductors, $g=-2e$, because a Cooper pair
consists of two electrons.)
The Lagrangian is invariant under
local (position-dependent) rotations of the phase angle
\begin{equation}
\phi\rightarrow e^{i\alpha}\phi,
\qquad
A_\mu \rightarrow A_\mu - \frac{1}{g}\partial_\mu\alpha.
\end{equation}
Again, we write $\phi=\phi_0+\delta\phi$, where we now choose $\phi_0$ to be 
real and positive. 
The gauge field $A_\mu$ does not have a homogeneous mode, and therefore
we can treat is as a pure fluctuation.
Again, we average the Euler-Lagrange equation for the real part $\phi_1$
over fluctuations
$\delta\phi_1$, $\delta\phi_2$ and $A_\mu$,
\begin{eqnarray}
0 &=&
\left\langle\frac{\partial{\cal L}}{\partial\phi_1}\right\rangle
\nonumber\\&=&
\lambda
\left(
\langle\phi_1^3\rangle
+
\langle\phi_2^2\phi_1\rangle - v^2\langle\phi_1\rangle
\right)+g^2\langle A_\mu A^\mu \phi_1\rangle
\nonumber\\&=&
\lambda\left[
\phi_0^2+3\langle\delta\phi_1^2\rangle
+\langle\delta\phi_2^2\rangle
+\frac{g^2}{\lambda}\langle A_\mu A^\mu\rangle
-v^2
\right]\phi_0.
\end{eqnarray}
%
The gauge field contains three physical degrees of freedom, and therefore
we have
\begin{equation}
\langle A_\mu A^\mu \rangle
=3\langle \delta\phi^2 \rangle=
\frac{T^2}{4} - \frac{3m_\gamma T}{4\pi} + (\mbox{$T=0$ part}).
\end{equation}
Crucially, however, the mass $m_\gamma$ of the gauge field depends on $\phi_0$
because of the Higgs mechanism, and is $m_\gamma=g\phi_0$. Including this
contribution, we find
\begin{equation}
\lambda\left[\phi_0^2
-\frac{3}{4\pi}\frac{g^3}{\lambda}T\phi_0
+\left(\frac{1}{3}+\frac{g^2}{4\lambda}\right)T^2-v^2\right]\phi_0=0.
\end{equation}
%
To interpret this results, we define the effective potential as the
tree-level potential that would give this same Euler-Lagrange equation.
That is obtained by simply integrating the left hand side of the equation,
\begin{eqnarray}
V_{\rm eff}(\phi_0) &=&
\frac{1}{2}\left[ \left(\frac{\lambda}{3} + \frac{g^2}{4}\right) T^2 
-\lambda v^2\right]\phi_0^2
\nonumber\\&&
-\frac{g^3T}{4\pi}\phi_0^3+\frac{\lambda}{4}\phi_0^4.
\label{equ:cubicterm}
\end{eqnarray}
The presence of the cubic term means that as long as
the coefficient of the quadratic term is positive and small enough, there
are two non-negative minima, 
one at $\phi_0=0$ and the other at a non-zero value.
The two minima are degenerate at a certain critical value of $T_c$,
\begin{equation}
T_c=\sqrt{\frac{12\lambda v^2}{4\lambda +3 g^2 - (3/2\pi^2)(g^6/\lambda)}}.
\end{equation}
This corresponds to a first-order transition line. Thus we see that
due to thermal fluctuations, we have a first-order transition instead of
a second-order one.\footnote{Note that this is the reason why Type I 
superconductors have a first-order transition.}
The discontinuity in $\phi_0$ is
\begin{equation}
\Delta\phi_0=\frac{1}{2\pi}\frac{g^3}{\lambda}T_c.
\label{equ:phidisc}
\end{equation}
It is also straightforward to calculate the range of temperatures in which a
second, metastable minimum exists,
\begin{equation}
\frac{12\lambda v^2}{4\lambda + 3 g^2}
< T^2 <
\frac{12\lambda v^2}{4\lambda +3 g^2 - (27/16\pi^2)(g^6/\lambda)}
\end{equation}
The critical temperature $T_c$, together with the metastability range
is shown in Fig.~\ref{fig:phasediag} for $g=0.5$. 
Fig.~\ref{fig:potential} shows the shape of the effective potential at $T_c$
and at the ends of the metastability range for $g=0.5$, $\lambda=0.05$.

When deriving Eq.~(\ref{equ:cubicterm}), we have assumed that the fluctuations 
are Gaussian. It is important to check whether this approximation, which is
equivalent to a one-loop calculation in terms of Feynman diagrams, is
consistent.
At higher orders in the loop expansion, one finds that each loop
gives typically a contribution of the order $g^2T/m_\gamma$.
At $T_c$, the photon mass  in the broken phase is
\begin{equation}
m_\gamma = g\Delta\phi_0=\frac{1}{2\pi}\frac{g^4}{\lambda}T_c,
\end{equation}
and therefore each loop order gives a contribution proportional to
$\lambda/g^2$. If $\lambda\ll g^2$, this is small and means that the 
contribution from higher loop diagrams can be ignored. One can
therefore trust the calculation of the effective potential in 
Eq.~(\ref{equ:cubicterm}). 
On the other hand, if $\lambda\agt g^2$,
the higher loop orders are not suppressed relative to the one-loop
contribution, and the whole approximation breaks down.

In the electroweak theory the ratio $\lambda/g^2$ is essentially
given by the ratio of the Higgs and $W$ boson masses. The Higgs
boson has not been found and its mass is therefore unknown, but
there is an experimental limit that its mass has to be at least 115 GeV. 
This means that $\lambda > g^2$, and the perturbative calculation breaks down.
This is also the case in Type II superconductors. In both cases,
one needs non-perturbative methods to study the properties of the
phase transition.

\section{Non-perturbative results}

In the global theory, one can be certain that no matter what 
non-perturbative
effects there may be near the transition, the transition must
exist, because one can measure the value of $\langle\phi\rangle=\phi_0$.
It vanishes in the symmetric phase and is non-zero in the broken phase,
and a function that does that cannot be analytic. In other words, 
$\langle\phi\rangle$ acts as an order parameter.

When one considers non-perturbative effects in the gauge theory, one makes
the simple but rather striking observation that $\langle\phi\rangle$ cannot
be an order parameter, because it is not physically observable~\cite{Elitzur}. 
This is because
it transforms non-trivially under gauge transformations,
\begin{equation}
\langle \phi(x)\rangle \rightarrow e^{i\alpha(x)}\langle \phi(x)\rangle.
\end{equation} 
If one averages
over all gauge-equivalent configurations, the expectation value of
any such quantity vanishes.

The most reliable way of studying phase transitions non-perturbatively is
to use lattice Monte Carlo simulations~\cite{Smit}, 
and they show that the theory
in Eq.~(\ref{equ:gaugeLag}) still has a transition. At small $\lambda/g^2$,
the transition is of first order as our perturbative calculation
indicated, but when $\lambda/g^2$ is larger, it becomes a continuous
second-order transition. Even though one cannot use $\langle\phi\rangle$
as an order parameter, one can see which phase the system is in by
studying the properties of the magnetic field. It is massless in
the symmetric phase and massive in the broken phase~\cite{Peisa}. 
The latter is 
equivalent to the Meissner effect in superconductors.

In non-Abelian theories, there are generally no massless fields, because
the theories are confining. Therefore there is no qualitative difference
between the ``symmetric'' and ``broken'' phases of the electroweak
theory. Careful lattice Monte Carlo simulations have shown that 
when the Higgs mass is above 80 GeV, the two phases are
actually smoothly connected and there is no phase transition between
them~\cite{Kajantie}. 
Because the true Higgs mass has to be in this range, it seems likely
that there was never a true electroweak phase transition
in the early universe. On the other hand, this conclusion may still change
if the Higgs sector of the electroweak theory is more complicated than
in the simplest version of the theory.

Similarly, the GUT phase transition may have been a smooth crossover from
one phase to another, provided that $\lambda/g^2$ was large.
This is perhaps surprising, because the broken phase of any GUT is supposed
to be the SU(3)$\times$SU(2)$\times$U(1) phase of the Standard Model, and
it contains a massless photon. However, the photon mass is not exactly zero
in GUTs~\cite{Polyakov}. 
It is finite, but suppressed by $\exp(-T_{\rm GUT}/T)$,
and therefore the photon is massless for all practical purposes. Nevertheless, 
an arbitrarily small mass is enough to make a smooth crossover possible.

\section{Conclusions}
We have seen that symmetries that are broken in
the vacuum are typically restored at high temperatures, 
and that this leads to phase 
transitions. In global theories, the transitions are generally of
second order, and in gauge theories they are of first order if 
the scalar self coupling $\lambda$ is small relative to the gauge coupling 
$g^2$. However, in the opposite limit, the transition may become 
a smooth crossover between two phases, analogously to the transition
between water and vapour at very high pressures.

At first sight, it seems there have been at least
three phase transitions related to particle physics in
the early universe: the deconfinement transition of QCD,
the electroweak transition and the GUT transition. On closer
scrutiny it turned out that it is likely that they may have all been
merely smooth
crossovers rather than true phase transitions in the mathematically rigorous
sense. Nevertheless, they may still have important consequences especially
in the form of defect formation.

\ 

\ 

\begin{center}
\bf\large 
Lecture II:\\
Defect Formation
\end{center}

\section{Introduction}
Topological defects are non-linear, time-independent solutions of 
field theories, which are made stable by the topology of the theory.
They have been studied extensively in cosmology~\cite{Vilenkin},
condensed matter physics~\cite{Volovik} and, at a more theoretical level,
in the context of quantum or classical field theory.

In cosmology, magnetic monopoles formed at the GUT phase transition
would have disastrous effects later on in the evolution of the universe,
and part of the motivation behind the inflationary theory~\cite{Guth:1980zm} 
was to wipe
them out before that. Cosmic strings, on the other hand, are fully
compatible with what is known about cosmology. If they were formed at
some early stage, we may well be able to detect them with astronomical
observations.

Topological defects also exist in many different condensed matter
systems. The ones that are closest to defects of cosmological importance 
are vortices in superfluids and Abrikosov flux lines in superconductors.

It was pointed out by Kibble in 1976~\cite{Kibble} that topological defects
are generally formed at symmetry breaking phase transitions and that we
should therefore expect that they were formed in the early universe, too.
However, as discussed in Ref.~\cite{proc1}, the known cosmological
phase transitions are not simple symmetry breaking transitions, but they
involve a breakdown of a local gauge symmetry. Nevertheless, as we
shall see, the same conclusion still holds: Topological defects are formed.

In recent years, 
defect formation has been studied extensively in condensed matter
systems such as liquid crystals~\cite{TurokNature,Bowick:1994rz,Digal:1998ak}, 
superfluids~\cite{Hendry,Dodd,Ruutu:1996qz,Bauerle} and 
superconductors~\cite{Carmi,Monaco,Kirtley,Maniv}.
The first two of these involve global symmetries, and indeed, they
seem to confirm Kibble's theory. Superconductors
have a U(1) gauge symmetry and would therefore be closer to the
cosmological phase transitions. The superconductor experiments
carried out so far have not produced conclusive results, but
hopefully this will change soon.

In this lecture, I will first explain how topological defects form in 
transitions associated with a breakdown of a global symmetry. 
This is known as the Kibble-Zurek mechanism.
Then, I will 
discuss gauge symmetries and show that the mechanism is actually rather
different in that case.
For a more comprehensive review, see Refs.~\cite{IJMPA,contphys}.

\section{Discrete global symmetries}

Consider first a real scalar field in a 
potential with a ${\mathbb Z}_2$ symmetry,
\begin{equation}
V(\phi)=\frac{1}{2}m^2\phi^2+\frac{1}{4}\lambda\phi^4.
\label{equ:globPot}
\end{equation}
If $m^2$ is positive, the only minimum is at $\phi=0$, and the state
of the system is symmetries under ${\mathbb Z}_2$. However, for $m^2<0$,
there are two degenerate minima at $\phi=\pm \sqrt{-m^2/\lambda}$.

If the field is initially exactly zero, it remains zero even if $m^2$ is 
negative. 
However, in real systems there are always both quantum and thermal 
fluctuations. Although the arguments we will use in this section are
probably also valid for pure quantum fluctuations, we assume that
the temperature is non-zero and 
the dominant
fluctuations are thermal. Therefore, we will not consider any quantum
effects.

If either the value of $\phi$ or its velocity is initially non-zero, 
it determines which minimum the system ends up in. 
If the initial condition is inhomogeneous, i.e., position-dependent,
different parts of the system may end up in different vacua.
On the whole, each minimum is equally probable, and therefore the system ends
up in a configuration of positive and negative regions, which both cover
half of the whole system. The boundaries between these domains are
topological defects
known as domain walls. The system is in vacuum on both sides of the wall,
but because of continuity, the field must leave the vacuum inside the
wall to interpolate between the positive and negative values.
This means that the wall has positive energy per unit area.

The most obvious question about the domain configuration is whether it
has a characteristic length scale. In general, the initial fluctuations
have a finite correlation length $\xi$, which is defined by
\begin{equation}
\langle \phi(\vec{x})\phi(\vec{y})\rangle
-\langle \phi(\vec{x})\rangle\langle\phi(\vec{y})\rangle
\sim e^{-|\vec{x}-\vec{y}|/\xi}.
\end{equation}
This means that if the points $\vec{x}$ and $\vec{y}$ are further away 
from each other than $\xi$, the initial conditions at these points are
independent. If we then assume that the domains form in a short enough time
that no signal can travel from $\vec{x}$ to $\vec{y}$ before they
have formed, there is a 50\% chance that the system will end up in
different vacua at these two points, in which case there would have to
be (at least) one domain wall between them.
Thus, one expects the typical distance between the walls to be roughly 
$\xi$.

Let us now assume that $m^2$ is initially positive but
decreases with time. Initially, the system is in thermal equilibrium
in the symmetric phase, but when $m^2$ reaches a critical value,
a symmetry breaking phase transition takes place.
Because the leading effect of a non-zero temperature is to change the
effective mass term~\cite{proc1}, we can equally well assume that $m^2$
is constant, but $T$ decreases due to, say, the expansion of the universe
or a contact with a colder heat bath.

It is a general property of second-order phase transitions that
the correlation length diverges at the transition point. This divergence
can be characterised by a critical exponent $\nu$, which is universal
in the sense that it only depends on the universality class and not
on any microscopic details,
\begin{equation}
\xi(T)=\xi_0\left(\frac{T_c-T}{T_c}\right)^{-\nu}.
\end{equation}
Typical values of $\nu$ are around 2/3.

If we could cool the system adiabatically so that it stays in equilibrium, 
the correlation length $\xi$ 
would be infinite at the transition temperature $T_c$. Consequently,
the whole system could end up in the same vacuum and there would be
no domain walls.

However, this is not possible in practice, because of critical slowing 
down~\cite{Zurek:qw}.
The relaxation time $\tau$, i.e., the time it takes for a small perturbation
to equilibrate, also diverges exponentially, with exponent $\mu$,
\begin{equation}
\tau(T)=\tau_0\left(\frac{T_c-T}{T_c}\right)^{-\mu}.
\end{equation}
This means that if we are cooling the system with a constant rate characterised
by the ``quench time'' $\tau_Q$,
\begin{equation}
T(t)=\left( 1 - \frac{t}{\tau_Q} \right) T_c,
\end{equation}
the equilibrium correlation length and relaxation time diverge as
\begin{equation}
\xi(t)=\xi_0\left(\frac{|t|}{\tau_Q}\right)^{-\nu},
\qquad
\tau(t)=\tau_0\left(\frac{|t|}{\tau_Q}\right)^{-\mu}.
\end{equation}
Once $\tau(t)>|t|$, any deviation from equilibrium will not
have time to disappear before the transition. The fluctuations in the
system at that time freeze out and survive all the way to the transition.
In particular this means that the correlation length cannot grow
significantly after that time, because it is simply a property of these
fluctuations.

This means that the actual correlation length at the time of the transition
is
\begin{equation}
\hat\xi = \xi(\hat t),
\end{equation}
where $\hat t$ is the freeze-out time determined by the equation
$\tau(\hat t)=|\hat t|$. This is straightforward to solve and gives
\begin{equation}
\hat t = - (\tau_0\tau_Q^\mu)^{1/(1+\mu)},
\end{equation}
whereby the frozen-out correlation length is
\begin{equation}
\hat\xi = \xi_0 \left(\frac{\tau_Q}{\tau_0} \right)^{\nu/(1+\mu)}
\propto \tau_Q^{\nu/(1+\mu)}.
\end{equation}
As we saw before, this is the typical distance between domain walls formed 
in the transition.

\section{Continuous global symmetries}
It is straightforward to generalise these arguments to systems with 
continuous symmetries. In the case of a complex scalar field with a
U(1) symmetry, the set of possible vacua, i.e., the vacuum manifold,
is a circle. The vacua are characterised by the phase angle $\varphi$
of the field $\phi=ve^{i\varphi}$.
After the transition,
the angle $\varphi$ is uncorrelated
at distances longer than $\hat\xi$.

\begin{figure}
\begin{center}
\epsfig{file=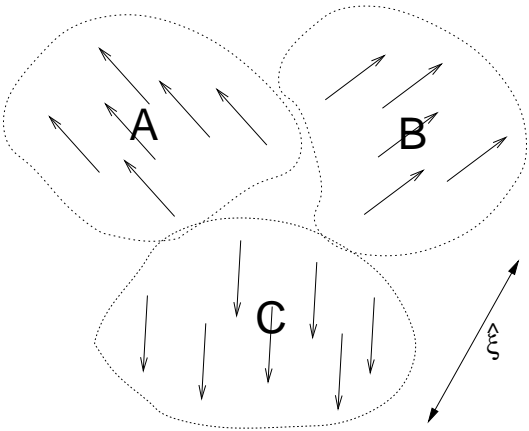,width=7cm}
\end{center}
\caption{\label{fig:domains}
The Kibble-Zurek mechanism is based on the observation that the choice of
the vacuum at points A, B and C separated by more than the correlation
length $\xi$ must be uncorrelated.
}
\end{figure}

Consider now three points A, B and C separated by more than $\hat\xi$ 
(see Fig.~\ref{fig:domains}). Each 
one has a random value of $\varphi$, but field $\phi$ is still continuous.
It will therefore have to somehow interpolate between the three points, and
if the points are not separated by much more than $\hat\xi$, it is most likely
that the field follows the shortest path on the vacuum manifold, because that
costs least energy. This assumption is known as the geodesic rule. However, 
there is a finite probability that if the geodesic rule is satisfied between 
A and B, between B and C and between C and A, the phase angle changes by 
$2\pi$ as one follows it from A to B to C to A. In that case, continuity
tell us that the field has to leave the vacuum manifold and vanish
at some point
within the circle formed by these points. That point is a vortex.

As with domain walls, the typical distance between vortices after the
transition is $\hat\xi$, and their number density is therefore
\begin{equation}
n\approx \hat\xi^{-2} \propto \tau_Q^{-2\nu/(1+\mu)}.
\label{equ:vortexscaling}
\end{equation}

It is straightforward to generalise these arguments to more complicated
symmetry groups. In particular, one finds that the number density of 
magnetic monopoles is $n\approx \hat\xi^{-3}$.

In principle, 
Eq.~(\ref{equ:vortexscaling}) makes a prediction that can be tested
in condensed matter experiments. However, the arguments that led to
it neglected factors or order one in many places, and therefore even if one
knew exactly the values of the parameters $\xi_0$ and $\tau_0$,
the prediction could still easily be off by a factor of ten. A more
robust prediction is the dependence on $\tau_Q$, but 
that would require carrying out experiments with a wide range of 
different cooling rates, and
in many cases it is
not possible to vary the cooling rate at all.

An even more robust test of the Kibble-Zurek mechanism is given by the spatial
distribution of vortices. Let us define $N_W(R)$ as the winding number
around a circle of radius $R$,
\begin{equation}
N_W(R)=\frac{1}{2\pi}\oint_{|\vec{r}|=R}d\vec{r}\cdot \vec{\nabla}\varphi.
\label{equ:NW}
\end{equation}
On average, this of course vanishes, but its 
variance $\langle N_W(R)^2\rangle$ is typically non-zero. After a phase 
transition in a global theory, the circle passes through 
around $2\pi R/\hat\xi$ domains that are uncorrelated with each other.
Every time, the curve moves from one domain to another, the phase angle
changes by some random amount $\Delta\varphi$. Labelling the domains by $i$, 
assuming the geodesic rule, 
and ignoring all factors of $2\pi$, 
the total winding number is then
\begin{equation}
N_W(R)\approx \sum_{i=1}^{R/\hat\xi}\Delta\varphi_i.
\end{equation}
Because the changes $\Delta\varphi_i$ are uncorrelated, the total variance
is
\begin{eqnarray}
\langle N_W(R)^2\rangle
&\approx&
\sum_{i,j}\langle \Delta\varphi_i\Delta\varphi_j\rangle
=\sum_i\langle \Delta\varphi_i^2\rangle
\nonumber\\
&=&\langle \Delta\varphi^2\rangle R/\hat\xi
\approx R/\hat\xi.
\end{eqnarray}
More generally, one can define a scaling exponent $\nu_D$ by
\cite{Digal}
\begin{equation}
\langle N_W(R)^2\rangle
\propto R^{4\nu_D}.
\end{equation}
The above result then implies that the Kibble-Zurek mechanism predicts
$\nu_D=1/4$.
This can be compared with a fully random distribution, in which the sign
$s_a$
of each vortex $a\in\{1,\ldots,N_{\rm vort}\}$
inside the curve is independent of all the others. In that case, the 
winding number is 
\begin{equation}
N_W(R)=\sum_{a=1}^{N_{\rm vort}} s_a,
\end{equation}
and consequently
\begin{eqnarray}
\langle N_W(R)^2\rangle
&\approx&
\sum_{a,b}\langle s_as_b\rangle
=\sum_a\langle s_a^2\rangle
\nonumber\\
&=&N_{\rm vort}\propto R^2.
\end{eqnarray}
Thus, a random distribution of vortices would have $\nu_D=1/2$.

\section{Gauge symmetries}
In a gauge field theory, one cannot apply the Kibble-Zurek arguments
as such, because the phase angle $\varphi$ is not a gauge invariant
quantity. Furthermore, while a global system tries to minimise phase
gradients $\vec{\nabla}\varphi$ in order to minimise energy, the
condition for minimum energy in a gauge theory is $\vec{\nabla}\varphi+
e\vec{A}=0$. If the gauge field $\vec{A}$ is non zero, the phase $\varphi$
does not even attempt to become uniform. 

In the special case in which the magnetic field is zero 
everywhere, one can choose a gauge in which $\vec{A}=0$. Then,
the gradients of $\varphi$ cost energy, and one can expect the
same arguments to hold as in a global theory.

However, if one considers a non-zero temperature, as we have been doing,
one cannot make this assumption, because it is not consistent with
thermal fluctuations~\cite{Hindmarsh:2000kd}. 
In fact, these fluctuations are nothing but
thermal radiation, and in the symmetric phase, they can be approximated
by the blackbody spectrum. To the extent that we can consider
classical dynamics, we can therefore say that the initial conditions
for the magnetic field are given by the Rayleigh-Jeans spectrum.
A convenient way to express that is by defining the function $G(k)$,
\begin{equation}
\langle B_i(\vec{k}) B_j(\vec{k}')\rangle
=
G(k)\left(\delta_{ij}-\frac{k_ik_j}{k^2}\right)
(2\pi)^3\delta(\vec{k}+\vec{k}').
\end{equation}
The Rayleigh-Jeans distribution corresponds to $G(k)=T$.

In the low-temperature phase, the long-wavelength part of the spectrum is
suppressed by the Higgs effect,
\begin{equation}
G(k)
=
\frac{Tk^2}{k^2+m_\gamma^2},
\label{equ:Gk}
\end{equation}
where $m_\gamma$ is the photon mass. These same considerations apply to 
superconductors, as well, and in that case $m_\gamma$ is the
inverse penetration depth.

Following Ref.~\cite{proc1}, one 
can calculate $m_\gamma$ as a function of temperature, and to leading
order one finds,
\begin{equation}
m_\gamma^2=e^2\phi_0^2=\frac{e^2}{4}(T_c^2-T^2).
\end{equation}
Assuming again linear cooling, one finds
\begin{equation}
m_\gamma^2=\frac{e^2T_c^2}{2}\frac{t}{\tau_Q}.
\label{equ:meanfieldmg}
\end{equation}

We will now assume that as the system is cooled through the transition,
the scalar field remains close to equilibrium.
This is not always true, and if it is not, then one would expect something
like the Kibble mechanism. However, here we are interested in the opposite
limit in which the gauge field falls out of equilibrium first.
We also assume that the mean field
result Eq.~(\ref{equ:meanfieldmg}) is approximately valid.

According to Eq.~(\ref{equ:Gk}), the amplitude $G(k)$ of a given mode
decreases slowly at first, but when $m_\gamma\approx k$, it starts to
drop sharply toward zero. Using Eq.~(\ref{equ:meanfieldmg}), we see
that this happens at time
\begin{equation}
t\approx t_k=\frac{2k^2}{e^2T_c^2}\tau_Q.
\end{equation}
After this time, the equilibrium value of $G(k)$ decreases as
\begin{equation}
G_{\rm eq}(k)\approx \frac{2k^2}{e^2T_c}\frac{\tau_Q}{t},
\end{equation}
and we find
\begin{equation}
\frac{d \ln G_{\rm eq}(k)}{dt}=-\frac{1}{t}.
\end{equation}
This rate is faster at early times, and the fastest 
decay takes place at $t\approx t_k$. Thus,
\begin{equation}
\left.\frac{d \ln G_{\rm eq}(k)}{dt}
\right|_{\rm max}\approx-\frac{1}{t_k}
=-\frac{e^2T_c^2}{2k^2\tau_Q}.
\label{equ:maxdecayrate}
\end{equation}
This is the rate at which $G(k)$ has to be able to decay for the mode
to stay in equilibrium. We can see that long wavelengths (low $k$) have
to decay faster.

On the other hand, one would expect on physical grounds that the dynamics
of longer wavelengths is slower and that they would therefore actually
decay shower than short wavelengths. It is therefore inevitable that some
very long-wavelength modes are unable to reach the necessary rate 
in Eq.~(\ref{equ:maxdecayrate}) and fall out of equilibrium instead.

We can be more specific if we assume that the dynamics is governed by the
conductivity $\sigma$ and Ohm's law is valid,
\begin{equation}
\vec{\jmath}=\sigma\vec{E}.
\end{equation}
Maxwell's equations then imply
\begin{equation}
\ddot{\vec{B}}=-\vec{\nabla}\times\vec{\nabla}\times\vec{B}
+\vec{\nabla}\times\vec{\jmath}
=-\vec{\nabla}\times\vec{\nabla}\times\vec{B}
-\sigma\dot{\vec{B}}.
\end{equation}
In Fourier space (and ignoring vector indices), this becomes simply
\begin{equation}
\ddot{B}+\sigma\dot{B}+k^2 B=0,
\end{equation}
which is nothing but a damped harmonic oscillator.
If $k<\sigma/2$, it has an exponential solution
\begin{equation}
B\propto \exp\left(-\frac{k^2}{\sigma}t\right).
\end{equation}
For $G(k)$, this means that any deviation from the equilibrium value 
will decay no faster than at this rate,
\begin{equation}
\frac{d\ln G(k)}{dt}\gtrsim -\frac{2k^2}{\sigma}.
\end{equation}
Thus, we find that the critical value $k_c$, defined in such a way that modes 
with $k<k_c$ fall out of equilibrium, is given by
\begin{equation}
\frac{2k_c^2}{\sigma}
=\frac{e^2 T_c^2}{2k_c^2\tau_Q}
\quad
\Rightarrow
\quad
k_c\approx \left(\frac{e^2T_c^2\sigma}{4\tau_Q}
\right)^{1/4}.
\end{equation}
One can then roughly say that as $m_\gamma\rightarrow\infty$, modes with
$k>k_c$ decay to zero, whereas those with $k<k_c$ retain their original 
amplitude. Thus, we would have the spectrum
\begin{equation}
G(k)=\left\{
\begin{array}{ll}
T,&k<k_c,\\
0,&k>k_c.
\end{array}
\right.
\label{equ:frozenGk}
\end{equation}
What this means is that the magnetic field has not disappeared completely.
However, because the field cannot penetrate the broken, superconducting,
phase, the magnetic field must be confined into flux tubes, i.e., vortices.
Thus, we have seen that in the gauge theory, vortices are formed by 
the non-equilibrium dynamics of the gauge field. This phenomenon is
knows as flux trapping~\cite{IJMPA}, because these vortices originate in the 
thermal fluctuations of the magnetic field, which do not have time to 
decay and
are instead trapped in vortices.

In general, the winding number $N_W$ and the magnetic 
flux $\Phi=\int d^2\vec{S}\cdot\vec{B}$ are related by the flux quantum
$\Phi_0=2\pi/e$, and for the case of a circle of radius $R$ 
in Eq.~(\ref{equ:NW}), we have
\begin{equation}
N_W(R)=\Phi(R)/\Phi_0\approx e\Phi(R).
\end{equation}
On average, this is zero, but as in the global case, the variance is
non-zero and can be calculated from $G(k)$,
\begin{eqnarray}
\langle N_W(R)^2\rangle
&\approx& e^2\langle \Phi(R)^2\rangle\nonumber\\
&=&e^2\!\int^R\! d^2x d^2y \langle B_z(x)B_z(y)\rangle\nonumber\\
&=&e^2\!\int^R\! d^2x d^2y 
\int\! \frac{d^3k}{(2\pi)^3}
e^{i\vec{k}\cdot(\vec{x}-\vec{y})}
G(k).
\end{eqnarray}
Using Eq.~(\ref{equ:frozenGk}), this is
\begin{equation}
\langle N_W(R)^2\rangle
\approx
\left\{\begin{array}{ll}
e^2T_cR,&\mbox{if $R> 1/k_c$},\\
e^2T_ck_c^3R^4,&\mbox{if $R< 1/k_c$}.
\end{array}\right.
\label{equ:NW2}
\end{equation}
The scaling exponent $\nu_D$ is therefore
\begin{equation}
\nu_D=\left\{\begin{array}{ll}
1,&\mbox{at short distances},\\
1/4,&\mbox{at long distances}.
\end{array}\right.
\end{equation}
Unfortunately, the long-distance value $1/4$ is equal to what the 
Kibble-Zurek mechanism gives, and cannot therefore be used to distinguish
between the two theories. Because it depends on the dimensionality, 
the situation is different in two-dimensional models that have been
studied numerically~\cite{Hindmarsh:2000kd,Stephens:2001fv}.
Note, however, that a two-dimensional film in three-dimensional space,
the typical setup of actual experiments, behaves like a three-dimensional 
system in this sense~\cite{Kibble:2003wt}.

\begin{figure}
\begin{center}
\epsfig{file=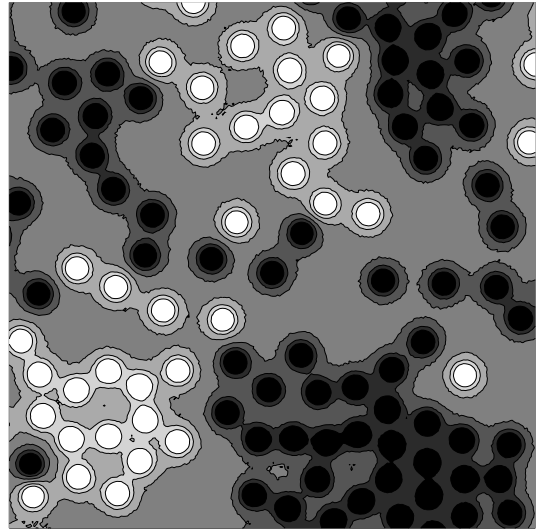,width=7cm}
\end{center}
\caption{
\label{fig:clusters}
A snapshot from a two-dimensional simulation by Stephens 
et al.~\cite{Stephens:2001fv} shows clearly the clusters
of positive (white) and negative (black) vortices formed in a
gauge field theory transition.
}
\end{figure}

In three dimensions, the best way to
distinguish between the theories is the short-distance exponent $\nu=1$.
It means that the typical value of $N_W(R)$ grows as $R^2$, i.e., it 
is proportional to the area of the circle. This is only possible if
all the vortices inside the circle have the same sign, or in other
words, if there are clusters of equal-sign 
vortices (see Fig.~\ref{fig:clusters}). 
Eq.~(\ref{equ:NW2}) tells us that the radius of these clusters 
is $\approx 1/k_c$, and the number of vortices in each 
is $\approx \sqrt{e^2T_c/k_c}$.
Consequently, the number density per unit cross sectional area is
\begin{equation}
n\approx\frac{\sqrt{e^2T_c/k_c}}{1/k_c^2}
\approx \sqrt{e^2T_ck_c^3}
\approx \sqrt{e^2T}
\left(
\frac{e^2T_c^2\sigma}{\tau_Q}
\right)^{3/8}.
\end{equation}

Here  we have only discussed Abelian gauge field theories, but similar
arguments seem to apply to non-Abelian theories as well. 
In that case, one can also study the formation of 
magnetic monopoles~\cite{Rajantie:2002dw}, which may shed more
light on the monopole problem in cosmology.

\section{Conclusions}
We have seen that if a symmetry breaking phase transition takes place in
a finite time, topological defects are formed.
In a global theory, they are produced by the Kibble-Zurek mechanism, which
leads to a characteristic negative correlation between vortices 
that can be used to
identify the mechanism. In a gauge theory, defects are also produced by
thermal fluctuations of the magnetic field that are too slow to decay.
The clearest prediction of this flux trapping scenario is the
formation of clusters of equal-sign vortices.

These theories are currently being tested in condensed matter experiments,
in particular with superconductors. Although the calculations done in
these paper refer to relativistic field theories, the qualitative conclusions
would be the same in condensed matter systems.

\

\begin{acknowledgments}
I would like to thank the organisers for a very successful 
workshop, and the ESF
COSLAB programme and Churchill College for financial support.
\end{acknowledgments}

\end{document}